\documentclass{PoS}

\title{$D$ to $K$ semileptonic decay form factors from HISQ light and charm quarks}

\ShortTitle{$D$ to $K$ semileptonic decays with HISQ action}

\author{\speaker{H.~Na}$^a$, C.~T.~H.~Davies$^b$, E.~Follana$^c$, P.~Lepage$^d$ and J.~Shigemitsu$^a$ \\ \\
   \llap{$^a$}Department of Physics, The Ohio State University, Columbus, Ohio, USA\\
   \llap{$^b$}Department of Physics and Astronomy, University of Glasgow, Glasgow, UK\\
   \llap{$^c$}Departamento de F\'{\i}sica Te\'orica, Universidad de Zaragoza, Zaragoza, Spain\\
   \llap{$^d$}LEPP, Cornell University, Ithaca, NY, USA\\ \\
        E-mail: \email{heena@mps.ohio-state.edu}}

\abstract{We present a study of $D \rightarrow K, l \nu$ semileptonic decays on the lattice which employs the HISQ action for both the charm and the light quarks. We work with MILC unquenched $N_f = 2 + 1$
lattices and determine the scalar form factor $f_0(q^2)$.
This form factor is obtained from a scalar current matrix element that does not require any operator matching.
We find $f^{D \rightarrow K}_0(0) \equiv f^{D \rightarrow K}_+(0) = 0.747(19)$
in the chiral plus
continuum limit and hereby improve the theory error on this quantity by a
 factor of $\sim$4 compared to previous lattice determinations. Combining the new
 theory result with  recent experimental
 measurements of the product $f^{D \rightarrow K}_+(0) * |V_{cs}| $
from BaBar and CLEO-c leads to a very precise
direct determination of the CKM matrix element
$|V_{cs}| $,
$|V_{cs}| = 0.961(11)(24)$, where the first error comes from experiment and the second
is the lattice QCD theory error.
}

\FullConference{The XXVIII International Symposium on Lattice Field Theory, Lattice2010\\
		June 14-19, 2010\\
		Villasimius, Italy}

\begin{document}

\section{Introduction}
From a study of $D \rightarrow K, l \nu$ semileptonic decays, one can calculate the form factor $f_+(q^2=0)$.
One can also determine the CKM matrix element, $|V_{cs}|$, by combining theory and experimental inputs.
We continue to work on the $D$ semileptonic decay project that was presented at the Lattice 2009 conference~\cite{lattice2009}.
In this article, we present a brief summary of our recent results for the $D$ to $K$ semileptonic decays, which are already published in Ref. ~\cite{prd2010}.
So, for more detail, please see the publication. 

For this project, we use $N_f=2+1$ asqtad MILC gauge configurations with two lattice spacings, $a \sim 0.12$fm ``coarse'' and $a \sim 0.09$fm ``fine'' ensembles. We apply the HISQ action for both the charm and light valence quarks. For better statistics, we employ random wall sources.
We develop a new extrapolation method to go to the continuum and chiral limit, the so called ``simultaneous modified $z$-expansion extrapolation,'' which allows us to extrapolate the form factors for the entire $q^2$ range.  This method does not have the expansion problem which normal chiral perturbation theory would have at large $E_K$.

To study the process $D \rightarrow K, l \nu$ one needs to evaluate the
matrix element of the charged electroweak current between the $D$ and the $K$ meson
states, $\langle K | (V^\mu - A^\mu) | D \rangle $.  Only the
vector current $V^\mu$ contributes to the pseudoscalar-to-pseudoscalar
amplitude and the matrix element can be written in terms of two
form factors $f_+(q^2)$ and $f_0(q^2)$, where $q^\mu = p^\mu_D - p^\mu_K$
is the four-momentum of the emitted W-boson.
\begin{eqnarray}
\label{vmu}
\langle K| V^\mu | D \rangle &=&  f^{D \rightarrow K}_+(q^2) \left [ p^\mu_D + p^\mu_K
- \frac{M^2_D - M^2_K}{q^2} \, q^\mu \right ] \\ \nonumber
 &+&  f^{D \rightarrow K}_0(q^2) \frac{M_D^2 - M_K^2}{q^2} q^\mu 
\end{eqnarray}
with $V^\mu \equiv \bar{\psi}_s \gamma^\mu \Psi_c$.
As described below, we find it useful to consider also the matrix element of
the scalar current $S \equiv \bar{\psi}_s \Psi_c$,
\begin{equation}
\langle K| S | D \rangle = \frac{M_D^2 - M_K^2}{m_{0c} - m_{0s}} f^{D \rightarrow K}_0(q^2).
\label{scalar}
\end{equation}
In continuum QCD one has the PCVC (partially conserved vector current)
relation and the vector and scalar currents obey,
\begin{equation}
q^\mu \langle V^{cont.}_\mu \rangle = (m_{0c} - m_{0s}) \langle S^{cont.} \rangle.
\label{pcvc}
\end{equation}
In fact PCVC is the reason why the same form factor $f^{D \rightarrow K}_0(q^2)$  appears
in eqs.(\ref{vmu}) and (\ref{scalar}).  On the lattice it is often
much more convenient to simulate with vector currents $\bar{\psi}_{Q1} \gamma^\mu 
\Psi_{Q2}$ that are not exactly conserved at finite lattice spacings even
for $Q1 = Q2$.
  Such non-exactly-conserved
currents need to be renormalized and acquire Z-factors.  We are able to carry out
fully nonperturbative renormalization of the lattice vector current by
imposing PCVC.  In the $D$ meson rest frame the condition becomes,
\begin{equation}
(M_D - E_K) \langle V_0^{latt.} \rangle Z_t +
\vec{p}_K \cdot \langle \vec{V}^{latt.} \rangle Z_s =
(m_{0c} - m_{0s}) \langle S^{latt.} \rangle.
\label{latpcvc}
\end{equation}

We have checked the feasibility of this renormalization scheme and
extracted preliminary
 $Z_t$ and $Z_s$ values for the test case of $D_s \rightarrow \eta_s, l \nu$
 in Ref.\cite{lattice2009}.  
However, here we focus on the form factor $f_+(q^2)$ just at
$q^2 = 0$, since this is all that is needed to extract $|V_{cs}|$.
We do this by exploiting  the kinematic identity $f_+(0) = f_0(0)$, and
concentrating on determining the scalar form factor $f_0(q^2)$ as accurately as
possible.  The best way to proceed is to evaluate the hadronic matrix
element of the scalar current rather than of the vector current.
From eq.(\ref{scalar}) one then has,
\begin{equation}
f^{D \rightarrow K}_0(q^2) =
 \frac{(m_{0c} - m_{0s}) \langle K | S | D \rangle}{M_D^2 - M_K^2}.
\label{f0}
\end{equation}
The numerator on the right-hand-side is a renormalization
group invariant combination.
This is true even in our lattice formulation, because we use the same relativistic action for both the heavy and the light valence quarks.
Moreover, eq.~(\ref{f0}) allows
a lattice determination of $f_0(q^2)$ and hence also of $f_+(0) = f_0(0)$
without any need for operator
matching.
Using eq.~(\ref{f0}) and going to the continuum limit is straightforward, because our action is so highly improved even for heavy quarks.
 
\section{Simultaneous modified $z$-expansion extrapolation}
The continuum $z$-expansion method is a well known model-independent parameterization method for semileptonic decay form factors. One can write the form factor as,
\begin{equation}
 f_0(q^2) = \frac{1}{P(q^2) \, \Phi_0(q^2,t_0)} \sum_{k = 0}^\infty
a_k(t_0) z(q^2,t_0)^k,
\label{f0zexp}
\end{equation}
where $P(q^2)$ and $\Phi_0(q^2,t_0)$ are given functions from analyticity properties of the form factors. 
 
\begin{figure}
\includegraphics*[width=7.0cm,height=8.0cm,angle=-90]{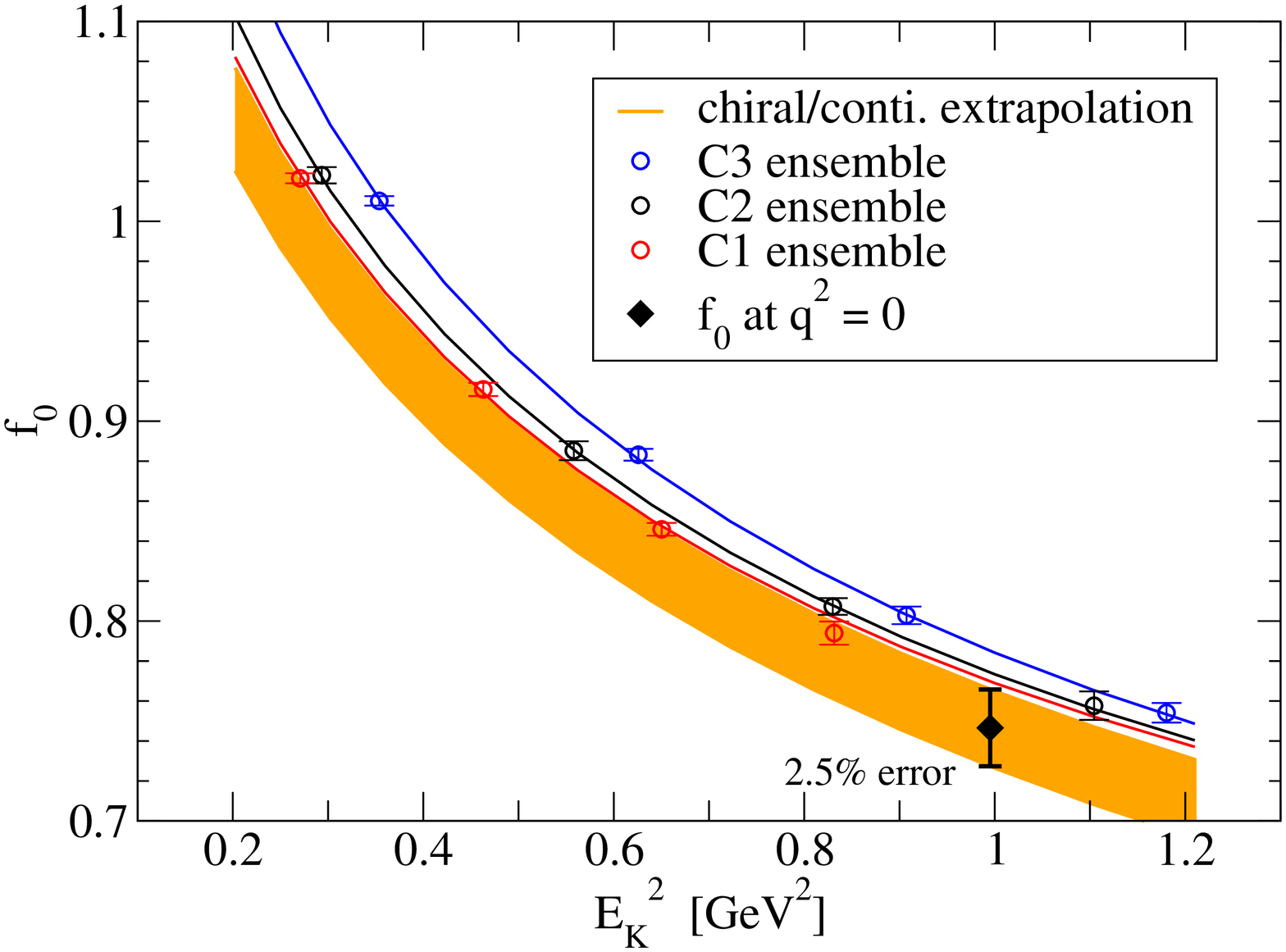}
\includegraphics*[width=7.0cm,height=8.0cm,angle=-90]{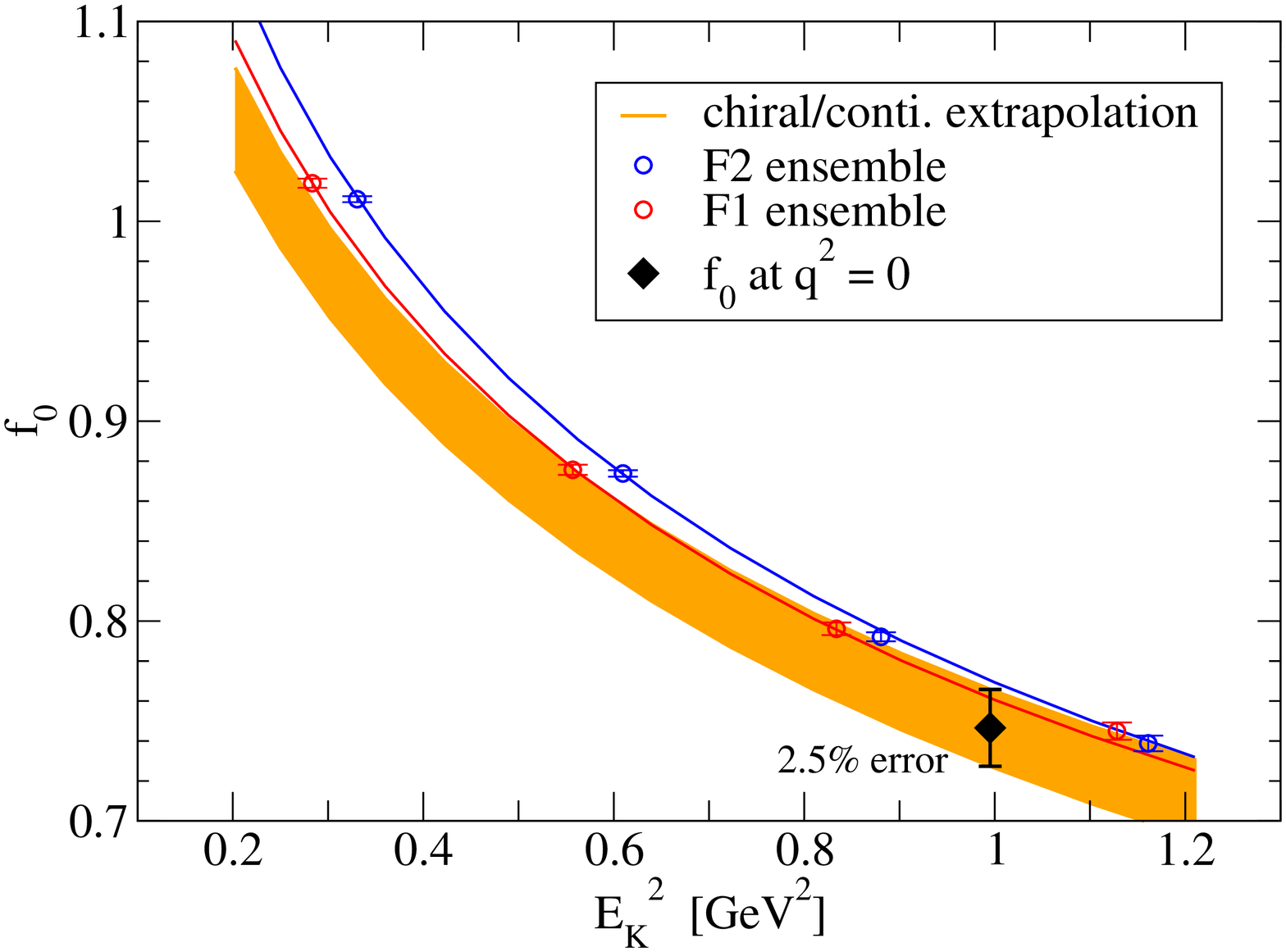}
\caption{
Chiral/continuum extrapolation of $f_0(q^2)$ versus $E_K^2$ from
the modified $z$-expansion ansatz.  The data points are coarse (left) and fine (right) lattice points.
Three individual curves and the extrapolated band are from a fit to all five ensembles. 
 }
\label{coarse1}
\end{figure}

\begin{figure}
\includegraphics*[width=7.0cm,height=8.0cm,angle=-90]{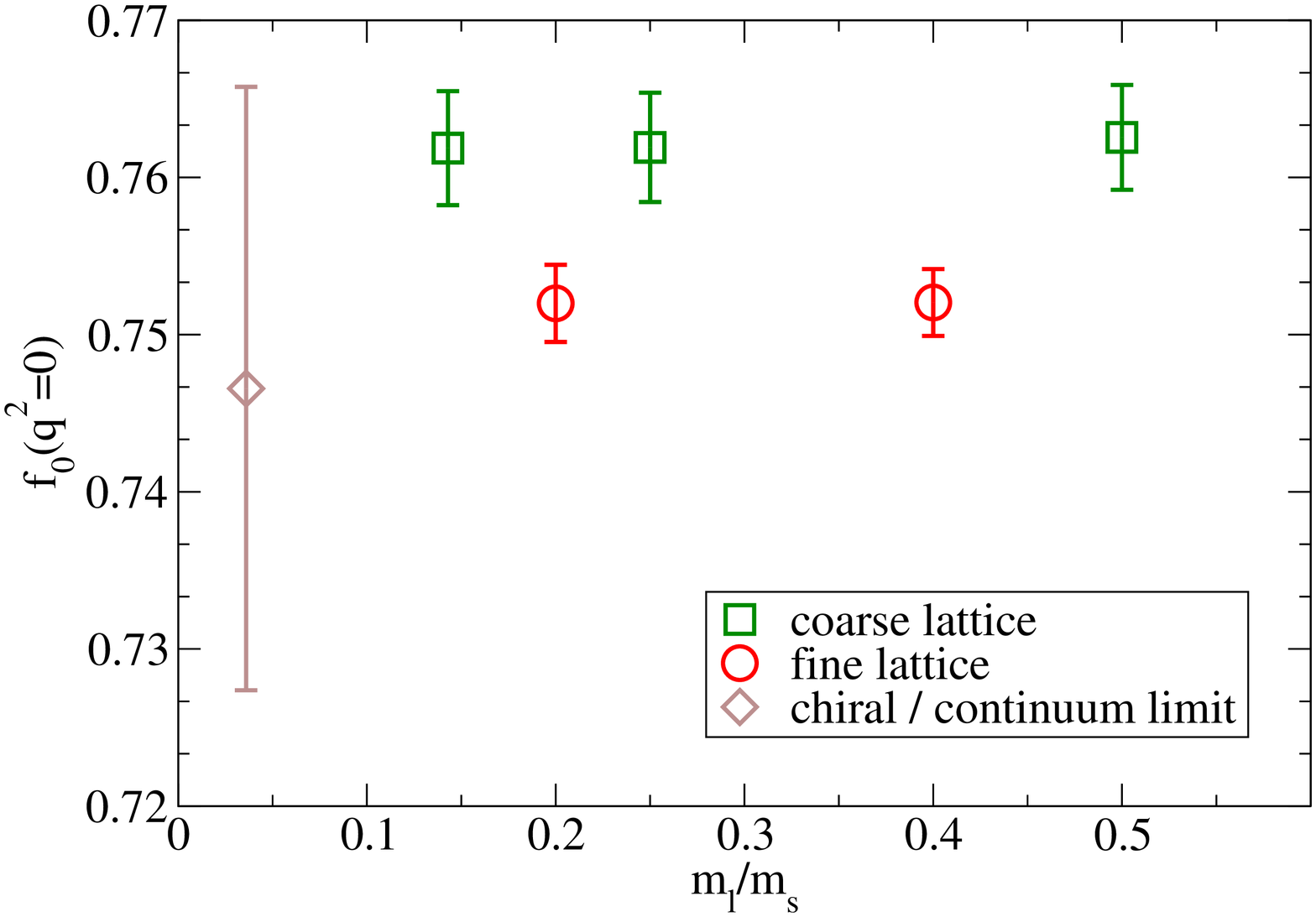}
\includegraphics*[width=7.0cm,height=8.0cm,angle=-90]{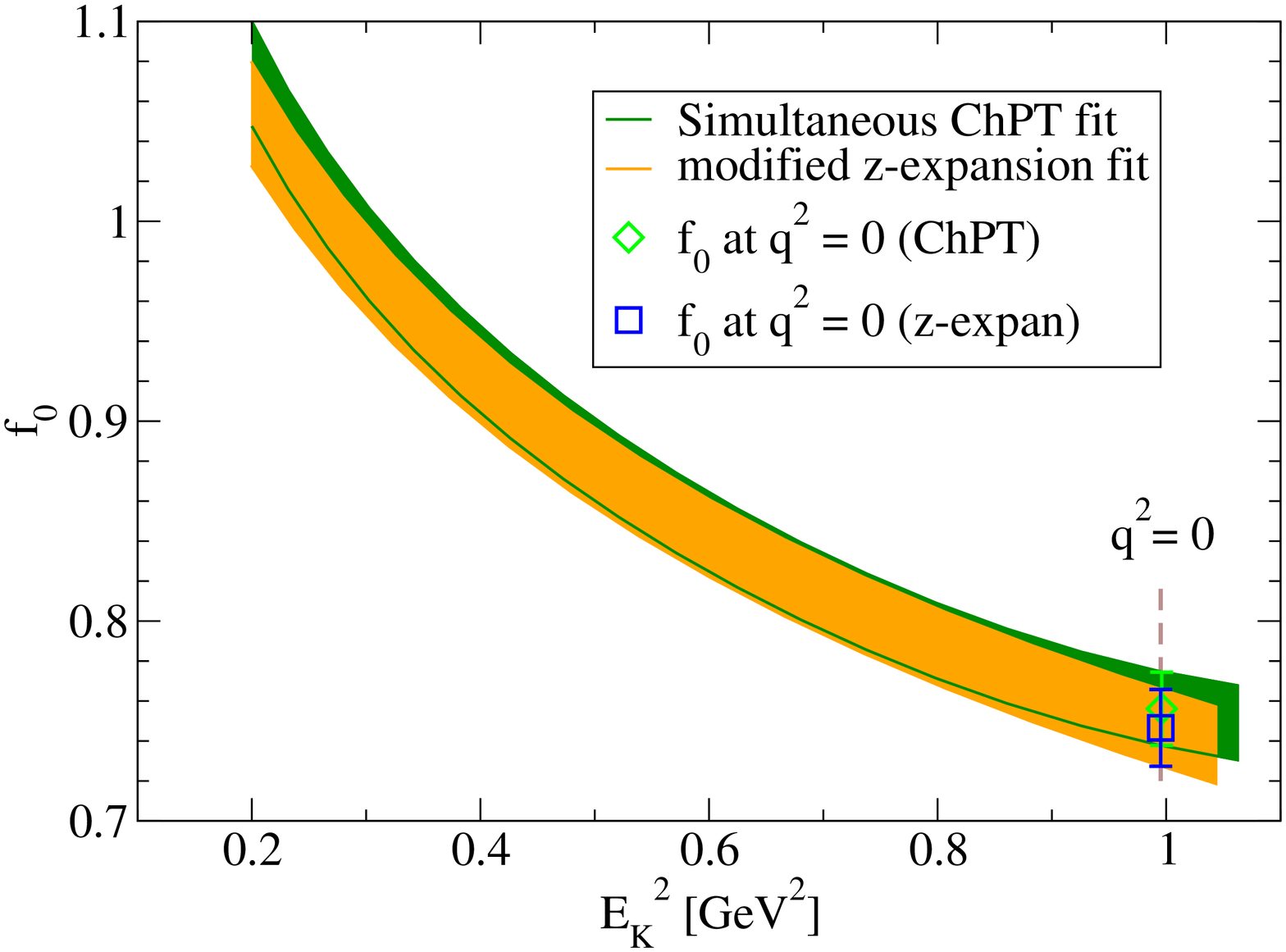}
\caption{
(left) $f_0$ at $q^2 = 0$ for the five ensembles and in the physical limit. (right) Comparisons of $f_0(q^2)$ in the physical limit from the $z$-expansion and
the ChPT extrapolations.
 }
\label{f0_q2}
\end{figure}

The $z$-expansion method works well for individual ensembles, however we like to modifying the fit ansatz to enable extrapolation to the physical limit.
All kinematic properties that depend on $q^2$ are absorbed by $P, \Phi_0,$ and $z$.
A natural way to distinguish between ensembles is to let $a_k \rightarrow a_k * D_k$, where $D_k$ contains the light quark mass and lattice spacing dependence as shown below with $k_{max} = 2$.
\begin{eqnarray}
f_0(q^2) &=& \frac{1}{P(q^2) \, \Phi_0} \left (a_0 D_0 + a_1 D_1 z + a_2 D_2 z^2 \right ) \\ \nonumber
& & \times (1+ b_1 (aE_K)^2 + b_2 (aE_K)^4),
\label{f0ansatz}
\end{eqnarray}
where,
\begin{eqnarray}
\label{di}
D_i &=& 1 + c^i_1 x_l + c^i_2 \delta x_s +c^i_3 x_llog(x_l) + d_i (am_c)^2 \\ \nonumber
   &&  + e_i (am_c)^4 + f_i \left ( \frac{1}{2} \delta M_\pi^2 + \delta M_K^2 \right ). 
\end{eqnarray}
In eq.~\ref{di}, we put typical analytic terms for light valence ($x_l$ and $\delta x_s$ terms) and sea quark mass ($\delta M_\pi$ and $\delta M_K$ terms) dependence.
For the chiral logs, we only include up/down quark contributions.
The strange quark chiral logs are close to a constant that can be absorbed into the $a_i$'s.
There are two distinct sources of lattice spacing dependence.
$(am_c)^2$ and $(am_c)^4$ terms are due to the heavy quark discretization error, and $(aE_K)^2$ and $(aE_K)^4$ terms are introduced to estimate the discretization errors due to finite momentum.
Since we want the $a_i D_i$ to be independent of the momentum, the $aE_K$ terms are placed separately outside the $z$-expansion.
We include lattice spacing dependent terms up to fourth power, however we tested with even higher terms and confirmed that the higher terms are negligible.
We have carried out simultaneous fits to all the data using the above ansatz and find
that very good fits are possible.   Fig.~\ref{coarse1} 
shows the resulting fit curves for each ensemble and the
chiral/continuum extrapolated
curve with its error band for $f_0(q^2)$ versus $E_K^2$
(we show separately the coarse and fine ensembles in order
to avoid too much clutter). 
On the left panel of Fig.~\ref{f0_q2} we show
$f_0(q^2=0)$ for the five ensembles and in the physical limit. One sees that
 within errors  this
quantity shows little light quark mass dependence and a $\sim 1.3 \%$ lattice
 spacing dependence.
We also test the chiral/continuum extrapolation with partially quenched chiral perturbation theory (PQChPT). This traditional method gives results in very good agreement with the modified $z$-expansion extrapolation method (see the right panel of Fig.~\ref{f0_q2}).

\section{$f_+(0)$, $|V_{cs}|$, and unitarity tests}
\subsection{$f_+(0)=f_0(0)$}
From the simultaneous modified $z$-expansion extrapolation method, we find $f_+(0) = 0.748 \pm 0.019$ in the physical limit for $D^0 
\rightarrow K^- l\nu$, and $f_+(0) = 0.746 \pm 0.019$ for $D^+ \rightarrow \overline{K}^0 l\nu$.
We take an average over these two channels and our final result in the physical limit becomes,
\begin{equation}
f^{D \rightarrow K}_+(0) = 0.747 \pm 0.011 \pm 0.015.
\label{resf+}
\end{equation}
The first error comes from statistics and the second error represents systematic errors.
Table~\ref{T.error} summarizes the error budget.
One sees that the largest contributions to the total error come from statistics
followed by $(am_c)$ and $(aE_K)$ extrapolation errors.

In order to calculate the form factor, we have to put in meson masses from experiment and also from our lattice simulations.
For example, we need experimental $D$, $K$, and $\pi$ meson masses to get the form factor at the physical limit, and $E_K$, $D$, and $K$ meson masses from the lattice calculations are used to fit at non-zero lattice spacing.
In Table~\ref{T.error}, ``Input meson mass'' refers to errors induced from these input meson masses.
In the fit ansatz, eq.~\ref{di}, there are light quark ($c_1^i$ and $c_3^i$), strange quark ($c_2^i$), and sea quark dependent terms ($f_i$).
Each systematic error due to these terms is shown on the fourth to sixth line in the table.
Lattice spacing dependence errors are estimated separately for $(am_c)^n$ and $(aE_K)^j$ type contributions.

In the fit ansatz, $x_l log(x_l)$ is the most infrared sensitive term.
We calculate the pion-tadpole loop integral both at finite volume and at infinite volume and compare these to estimate the finite volume effects.
For the charm quark mass tuning error, we calculate the form factor with a different charm quark mass, $am_c = 0.629$, on the C3 ensemble, and compare with the result with the tuned $am_c = 0.6235$.

\begin{table}
\begin{center}
\begin{tabular}{|c|c|}
\hline
Type & Error \\
\hline
\hline
Statistical & 1.5 \% \\
Lattice scale ($r_1$ and $r_1/a$) & 0.2 \% \\
Input meson mass & 0.1 \% \\
Light quark dependence & 0.6 \% \\
Strange quark dependence & 0.7 \% \\
Sea quark dependence & 0.4 \% \\
$am_c$ extrapolation& 1.4 \% \\
$aE_K$ extrapolation& 1.0 \% \\
Finite volume & 0.01 \% \\
Charm quark tuning & 0.05 \% \\
\hline
Total & 2.5 \% \\
\hline
\end{tabular}
\end{center}
\caption{ Total error budget. }
\label{T.error}
\end{table}

\begin{figure}
\includegraphics*[width=7.0cm,height=8.0cm,angle=-90]{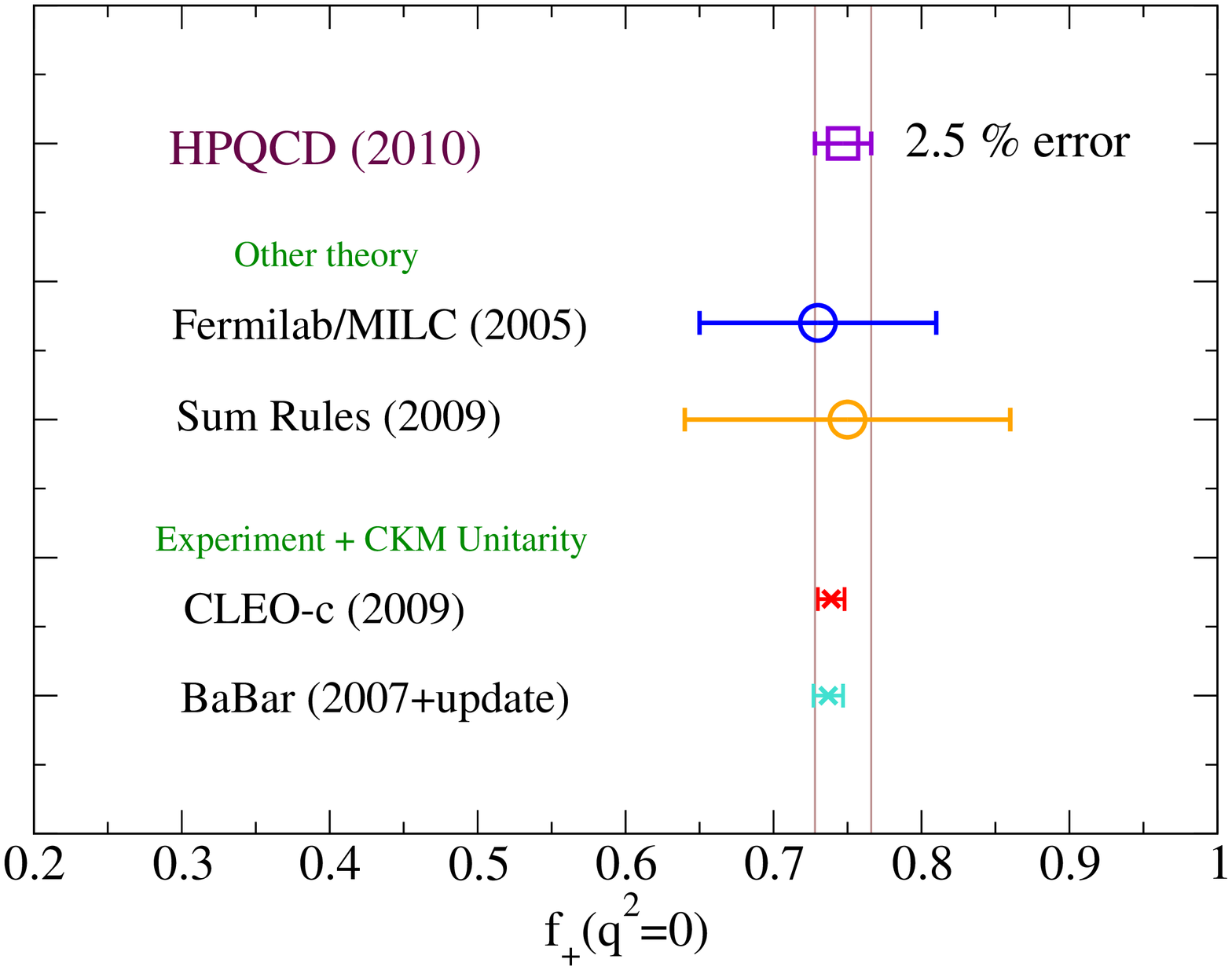}
\includegraphics*[width=7.0cm,height=8.0cm,angle=-90]{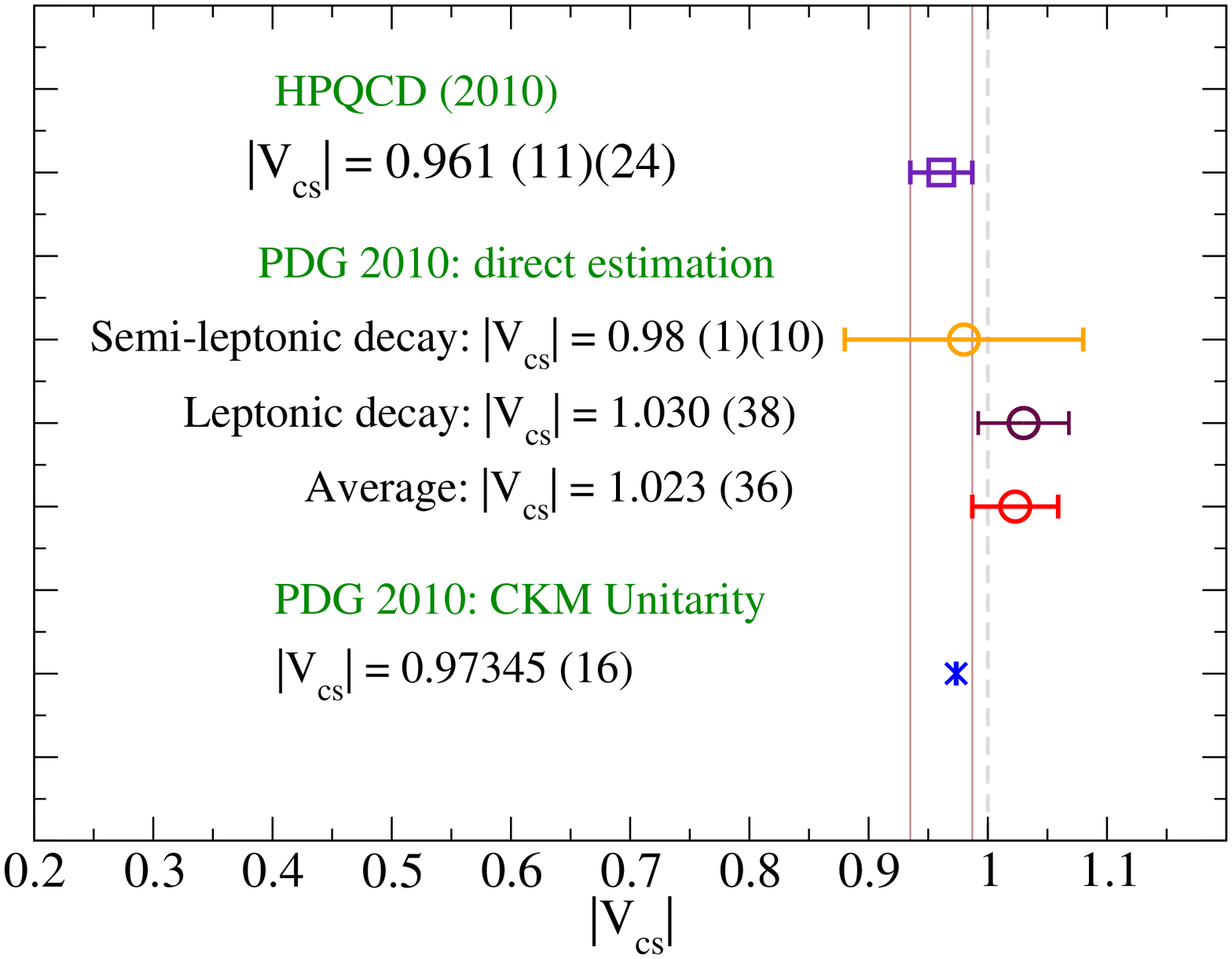}
\caption{
(left) Comparisons of $f_0(q^2=0)$ with other calculations and experiments.
 (right) Comparisons of our new $|V_{cs}|$ with values in the PDG \cite{pdg}.
 }
\label{resultf0}
\end{figure}

In their papers both BaBar~\cite{babar} and CLEO-c~\cite{cleo} have converted their measurements of $f_+(0) * |V_{cs}|$ into results
for $f_+(0)$ using values for $|V_{cs}|$ fixed by CKM unitarity.  For this CLEO-c  uses 
the 2008 PDG CKM unitarity value of $|V_{cs}| = 0.97334(23)$ and obtains
$f^{D \rightarrow K}_+(0) = 0.739(9)$
 and BaBar uses $|V_{cs}| = 0.9729(3)$ leading to $f_+(0) = 0.737(10)$.
On the left panel of Fig.~\ref{resultf0} we plot our result, eq.(\ref{resf+}), together with earlier
theory results from the lattice \cite{fermimilc05} and from a recent sum rules calculation and with the BaBar and CLEO-c numbers.
 One sees the very welcome reduction in theory errors which are now small enough so that the agreement between theory and experiment
 already provides a nontrivial indirect test of CKM unitarity.

\subsection{Direct Determination of $|V_{cs}|$ and unitarity tests}
As experimental input we take
$f_+(0) * |V_{cs}| = 0.719 (8)$ from CLEO-c \cite{cleo} and
$f_+(0) * |V_{cs}| = 0.717 (10)$ from BaBar \cite{babar}.  For the latter we have multiplied
BaBar's quoted $f_+(0)$ with their quoted CKM unitarity value for $|V_{cs}|$.
Averaging between the two experiments we use
$f_+(0) * |V_{cs}| = 0.718 (8)$ together with eq.(\ref{resf+}) to extract $|V_{cs}|$.
One finds,
\begin{equation}
|V_{cs}| = 0.961 \pm 0.011 \pm 0.024,
\label{resvcs}
\end{equation}
in good agreement with the CKM unitarity value of 0.97345(16) \cite{pdg}.
The first error in (\ref{resvcs}) is from experiment and the second from the lattice calculation of this article.
This is a very precise direct determination of $|V_{cs}|$,
made possible by the many advances in lattice QCD that are described in this article
together with the tremendous progress in recent experimental studies of $D$ semileptonic decays
\cite{babar,cleo}.
On the right panel of Fig.~\ref{resultf0} we plot several previous direct determinations of $|V_{cs}|$ from the 2010 PDG \cite{pdg}
together with (\ref{resvcs}) and the CKM unitarity value.

Using the new value of $|V_{cs}|$, eq.(\ref{resvcs}), and the current PDG  values
$|V_{cd}| = 0.230(11)$ and $|V_{cb}| = 0.0406(13)$ one finds,
\begin{equation}
|V_{cd}|^2 + |V_{cs}|^2 + |V_{cb}|^2 = 0.978(50)
\label{2ndrow}
\end{equation}
for the 2nd row. And similarly for the 2nd column, with $|V_{us}| = 0.2252(9)$ and $|V_{ts}= 0.0387(21)$
one gets,
\begin{equation}
|V_{us}|^2 + |V_{cs}|^2 + |V_{ts}|^2 = 0.976 (50).
\label{2ndcol}
\end{equation}

\section{Discussion}
We have carried out a successful calculation for $D \rightarrow K, l \nu$ semileptonic decay form factors using the HISQ action for both the charm and light quarks with $N_f=2+1$ asqtad MILC gauge configurations.
The total error for $f_+(0)$ is estimated here to be 2.5\%.  This is a factor of four times smaller than in the previous lattice calculation of Ref.~\cite{fermimilc05}.
This was achievable because of applying several new methods and techniques.
We employ the HISQ action for both charm and light quark actions and a scalar current rather than the traditional vector current.
Because of these new methods, we obtain results with smaller discretization errors and no operator matching.
We also developed the modified $z$-expansion extrapolation method, which is crucial to decrease errors due to the discretization, chiral / continuum extrapolation and parameterization of the form factor.
In order to decrease statistical errors, we apply random-wall sources and perform simultaneous fits with multiple correlators and $T$'s.
If we compare with the error budget of Ref.~\cite{fermimilc05}, then we see the statistical errors reduced from 3\% to 1.5\% and
the extrapolation and parameterization errors from 3\% to 1.5\% as well.
The biggest improvement is in the discretization errors. The total discretization errors have now been reduced from 9\% to 2\%.
We note that the concept of the discretization errors is different in Ref.~\cite{fermimilc05} compared to ours.
In Ref.~\cite{fermimilc05}, they estimate the discretization errors by power counting, since they calculate at only one lattice spacing.
Here, however, we actually perform continuum extrapolations with correction terms for the discretization effects.
As a result, we do not have discretization errors per se, but instead extrapolation errors due to higher order correction terms.

Again, this is a short version of Ref.~\cite{prd2010}. For more detail and full discussion, please see the publication.

\end{document}